\documentclass[twocolumn,showpacs]{revtex4}
\pdfoutput=1
\usepackage{graphicx}
\usepackage{amsmath}
\usepackage{amssymb,amsthm}
\graphicspath{{pict/}{}}

\usepackage{bm}

\newcounter{Fig}

\newcommand{\be}{\begin{equation}}
\newcommand{\ee}{\end{equation}}

\begin{document}
\title{Nonlinear localized modes in dipolar Bose-Einstein condensates in optical lattices}
\author{S. Rojas-Rojas$^{1,2}$, R. A. Vicencio$^{1,2}$, M. I. Molina$^{1,2}$, and F. Kh. Abdullaev$^{3,4}$}
\affiliation{$^{1}$Departamento de F\'{\i}sica, Facultad de Ciencias,
Universidad de Chile, Santiago, Chile\\
$^{2}$Center for Optics and Photonics (CEFOP), Casilla 4016, Concepci\'{o}n, Chile\\
$^{3}$Physical-Technical Institute, Uzbek Academy of Sciences, 2-b, G. Mavlyanov str., 100084, Tashkent, Uzbekistan\\
$^{4}$Centro de Fisica Teorica e Computacional, Universidade de Lisboa, Av. Prof. Gama Pinto 2, Lisboa 1649-003,Portugal}

\begin{abstract}
The modulational instability and discrete matter wave solitons in dipolar BEC, loaded into a deep optical lattice, are investigated analytically and numerically. The process of modulational instability of nonlinear plane matter waves in a dipolar nonlinear lattice is studied and the regions of instability are established. The existence and stability of bulk discrete solitons are analyzed analytically and confirmed by numerical simulations. In a marked contrast with the usual DNLS behavior (no dipolar interactions), we found a region where the two fundamental modes are simultaneously unstable allowing enhanced mobility across the lattice for large norm values. To study the existence and properties of surface discrete solitons, an analysis of the dimer configuration is performed. The properties of symmetric and antisymmetric modes including the stability diagrams and bifurcations are investigated in closed form. For the case of a bulk medium, properties of fundamental on-site and inter-site localized modes are analyzed. On-site and inter-site surface localized modes are studied finding that they do not exist when nonlocal interactions predominate with respect to local ones.

\end{abstract}

\maketitle

\section{Introduction}

The dynamics of Bose-Einstein condensates (BECs) in optical lattices has been the subject of intensive theoretical and experimental investigations in recent times~\cite{Morsh,Braz}. Different phenomena like generation of coherent packets of matter waves (atom lasers)~\cite{Kasevich}, Bloch oscillations~\cite{Morsh1,Carusotto,Salerno08}, gap solitons~\cite{Eier}, discrete breathers~\cite{Tromb01,Abd01}, compactons~\cite{Abd10} have been predicted and experimentally observed.

The possibility to vary different parameters of BEC systems - trap potential and atomic interactions - makes BEC a unique system for modeling different fundamental phenomena in condensed matter physics. The control of the strength and shape of the trapping potential induced by counter-propagating laser fields, as well as the time modulations of the parameters of the optical lattice, can be easily achieved.
The strength of atomic interactions and, thus, the mean field nonlinearity can also be tuned in space or time by using, for example, the Feshbach resonances method~\cite{Inouye}. This allows the atomic scattering length $a_s$ to be tuned by a time-or-space variation of the external magnetic field near the resonant value.

Joint effects of nonlinearity, periodicity and quantum pressure, lead to the existence of stable localized states conserving the form upon propagation and  collisions. Gap solitons, discrete breathers and discrete compactons are examples  of such states. Some of these structures are observed in experiments. Nonlinearity has usually been considered as local. Recent discovery of dipolar condensate with long-range interactions between atoms, posed the question of the existence of solitons in a dipolar BEC loaded in the optical lattice. The existence of solitons is due to the interplay between local (contact interactions) and nonlocal (long range interactions) nonlinearities and the optical lattice effects. Previously, the existence of discrete solitons in systems with nonlocal interactions has been studied for semiconductors amplifiers~\cite{Utanir} and nematic liquid crystals~\cite{Fratalocchi}. The first observation of a condensate in a gas of chromium atoms ($^{53}$Cr) with long-range interactions was reported in Ref.~\cite{1}. Dipolar condensate exhibits many unusual properties not encountered in BECs with local interactions, e.g. the existence of stable isotropic and anisotropic 2D solitons~\cite{2}.

Two limiting cases can be distinguished: shallow and deep optical lattices. Both cases have been considered recently~\cite{cuevas}. In the case of deep lattices, a discrete model using a nonlocal Gross-Pitaevskii equation is investigated in Ref.~\cite{4}. Here, the dynamics of unstaggered bright discrete solitons for a 2D {\it disk}-shaped dipolar BEC was analyzed. The system was studied by a 2D model based on the discrete nonlinear Schr\"{o}dinger (DNLS) equation, with on-site and long-range cubic nonlinearities. The existence of stable fundamental discrete solitons of the different symmetries was shown. The authors also observed a stability exchange of the fundamental solutions but did not study in detail the region where both solutions are simultaneously unstable. This is one the main goals of the present work, to explore in detail these regions and their dynamical properties. We will show that a very good mobility can be observed for solutions with a large value of the norm, what is absolutely forbidden in systems with only local nonlinearities.

It is currently of interest to investigate discrete breathers in quasi-one-dimensional {\it cigar}-shaped dipolar BEC loaded in a deep optical lattice. Recently, this problem was considered in Ref.~\cite{Malomed09}, where a non-polinomial DNLS model was obtained. The existence and stability of unstaggered bulk bright breathers was studied. Besides unstaggered bulk solitons, there is especial interest on staggered bulk solitons, existence of surface discrete breathers (solitons) and the modulational instability of matter waves in the dipolar BEC embedded in an optical lattice. Surface solitons are the generalization of nonlinear Tamm states, and they have been well-studied in DNLS systems with on-site nonlinearity~\cite{Tamm,surface}. The existence of nonlinear Tamm states in quantum dipolar gases in optical lattices is a fundamental problem. We can expect a rich variety of Tamm states due to the competition between local on site and nonlocal cubic interactions and the lattice potential. In the present work, we explore in detail this issue.

This paper is organized as follows: Section II introduces the discrete model of a Bose-Einstein condensate in a deep optical lattice subjected to dipolar interactions; in section III, we examine the modulational stability properties of nonlinear plane matter wave solutions; in section IV, we review some results for bulk localized modes, including some newly found bistable behavior that contrast markedly with bulk phenomenology found in usual DNLS systems. In section V, we focus on surface localized states and, finally, section VI concludes the paper.

\section{Lattice model}
The system under consideration is the quasi one-dimensional dipolar BEC in a deep optical lattice. The governing equation is the 1D Gross-Pitaevskii equation (GPE) with nonlocal interaction terms~\cite{5}:
\begin{eqnarray}
& & i\hbar {\partial \Psi\over{\partial T}} + {\hbar^2\over{2m}} {\partial^{2} \Psi\over{\partial X^{2}}} - V_0\cos(2kX) \Psi + g_{1D} |\Psi|^{2} \Psi  \nonumber\\
& &+ \frac{2\alpha d^2}{l_{\perp}^3} \Psi(x,t)\ \int_{-\infty}^{\infty} d\xi R(|X-\xi|)\ |\Psi(\xi,t)|^{2} = 0,\ \ \ \ \ \
\label{e1}
\end{eqnarray}
where $g_{1D} = 2\hbar a_s\omega_{\perp}$. $\omega_{\perp}$ corresponds to the transverse trap frequency, $l_{\perp} = \sqrt{\hbar/m\omega_{\perp}}$, and $d$ is the dipolar moment. The parameter $\alpha$ can vary from $1$ to $-1/2$ for dipoles oriented along or perpendicular to the $x$-axis. The wave function is normalized to the number of atoms comprising the BEC, $N\equiv\int_{-\infty}^{\infty} |\Psi(x)|^{2} dx$. Now we define dimensionless parameters:
\begin{eqnarray*}
E_R = \frac{\hbar^2 k^2}{2m}= \hbar\omega_R=\frac{V_0}{V},\ t = T \omega_R,\ x = k X,\\
g_0 = \frac{\alpha a_d}{l_{\perp}ka_{s0}},\ q_0 = \frac{a_s}{a_{s0}},\ \psi = \sqrt{\frac{2a_s \omega_{\perp}}{\omega_R}}\Psi,
\end{eqnarray*}
where $a_d = md_{d}^{2}/\hbar^2$ is the characteristic scale of the long-range dipolar interactions, and $a_{s0}$ is the background value of an atomic scattering length.
Equation (\ref{e1}) is recasted as:
\begin{eqnarray}
& & i\ {\partial\psi\over{\partial t}} + {1\over{2}} {\partial^{2}\psi\over{\partial x^{2}}} - V\cos(2x) \psi + q_0 |\psi|^{2} \psi  \nonumber\\
& &+ g_0\ \psi(x,t)\ \int_{-\infty}^{\infty} d\xi R(|x-\xi|)\ |\psi(\xi,t)|^{2} = 0\ \ \ \ \ \
\end{eqnarray}
where $\psi(x,t)$ represents the mean-field wave function of the condensate.
Two commonly used nonlocal kernel are,
\[
R_{1}(x) = (1 + 2 x^2)\exp(x^2) \mbox{erfc}(|x|) - 2 \pi^{-1/2} |x|,\ \text{and}
\]
\[
R_{2}(x) = \delta^{3} (x^{2}+ \delta^{2})^{-3/2}.
\]
The first kernel corresponds to a dipolar BEC in a quasi-1D trap~\cite{5}; while the second kernel, with the cutoff parameter $\delta$, is more convenient for an analytical treatment~\cite{6}. The matching condition
$R_{1}(0)=R_{2}(0)$ implies $\delta = \pi^{-1/2}$. Parameter $\delta$ corresponds to the effective size of the dipole. It takes a value of the order of the transverse confinement length, which makes the model one-dimensional, and constitutes the unit of  length in Eq.(\ref{e1}). Therefore, the choice of $\delta =\pi^{-1/2}\sim 0.56$ is quite reasonable. In the limit $x\gg \delta$, where dipole-dipole interaction effects dominate the contact interaction effects, both response functions behave as $\sim 1/x^{3}$.

In a deep optical lattice, $V \gg 1$ and it is natural to consider the expansion
\[
\psi = \sum_{n=-\infty}^{\infty} u_{n}(t)\ \phi_{n}(x),
\]
where $\phi_{n}(x)$ are Wannier-like functions located on the minima of the periodic potential $V(x)$. The analysis of overlap integrals \cite{Roati,3} shows that  the equations for $u_{n}$ become:
\begin{eqnarray}
i {d\over{dt}} u_{n}(t)&+&\kappa (u_{n+1}+u_{n-1}) + q |u_{n}|^{2} u_{n}+\nonumber\\
& &g (|u_{n+1}|^{2} + |u_{n-1}|^{2})u_{n} = 0,\label{eq:1}
\end{eqnarray}
where
\begin{eqnarray*}
q & = &q_0 \int\int R(x-\xi)|\phi_n(x)|^2|\phi_n(\xi)|^2 dx\ d\xi \ \ \text{and}\nonumber\\
g & = & g_0 \int\int R(x-\xi) |\phi_{n \pm 1}(x)|^2 |\phi_n(\xi)|^2 dx\ d\xi \ .
\end{eqnarray*}
It should be noticed that the parameter $q$, which is proportional to the atomic scattering length $a_s(t)$, can be set to zero by means of the Feshbach resonance method~\cite{Inouye}. According to this technique, by a variation of the external magnetic field near the resonant value, it is possible to diminish to zero the atomic scattering length responsible for the mean field nonlinearity parameter $q$. This limit of the model has been applied recently to the analysis of the fundamental limit on the atomic interferometer based on BEC with tunable scattering length, loaded in optical lattices~\cite{Roati}.

Equation (\ref{eq:1}) possesses two conserved quantities, the norm $N$ and the Hamiltonian $H$,
\begin{eqnarray}
N &=& \sum_{n=1}^{M} |u_{n}|^{2}\nonumber\\
H &=& -\sum_{n=1}^{M}\left[\kappa u_{n+1} u_{n}^{*}+ {q\over{4}} |u_{n}|^{4} + \frac{g}{2} |u_{n+1}|^{2} |u_{n}|^{2}+c.c. \right]\nonumber  \ ,
\end{eqnarray}
where $M$ indicates the number of lattice sites. Linear plane-wave solutions of Eq.(\ref{eq:1}) take the form $u_{n}(t)=u_{0} \exp(i k n + \omega t)$ and satisfy the linear dispersion relation $\omega=2 \kappa \cos k$, which defines the band of single-particle energies $\omega_k \in \{-2 \kappa, 2 \kappa\}$. Outside of this band, nonlinear localized solutions are expected to exist. $u_0$ and $k$ correspond to the normalized amplitude and quasimomentum of the condensate, respectively.

Along this work, we will look for bulk and surface one-dimensional fundamental, centered on-site and inter-site, solutions. By implementing a standard Newton-Raphson method, we numerically compute localized stationary solutions of model (\ref{eq:1}), of the form $u_n(t)=u_n \exp{[i\omega t]}$, by solving the set of equations
%
\begin{equation}
\omega\ u_{n}=(u_{n+1}+u_{n-1}) + q u_{n}^{3} + g(u_{n+1}^{2} + u_{n-1}^{2})u_{n}.\label{esta}
\end{equation}
%
where $u_n\in R$. Hereafter we will consider $\kappa=1$ and $q>0$. The regime $q <0$ can be explored by simply making the transformation $u_n\rightarrow (-1)^n u_n$,
$\omega\rightarrow -\omega$ and $g\rightarrow -g$. For each solution, we characterize it by computing its norm and Hamiltonian. We perform a linear stability analysis by using a standard method developed in Ref.~\cite{sta}. We obtain an eigenvalue spectrum and compute the eigenvalue with the largest imaginary part (denoted as $G$) as an indication of the instability gain. When plotting the norm- frequency diagrams of each mode (bulk or surface), we will use {\em continuous (dashed) lines to denote stable (unstable) solutions}. In addition, and as a visual aid, quantities of interest for the on-site (inter-site) solutions will be plotted in black (orange).

\section{Modulational stability}

We begin by studying the linear stability of plane waves solutions under the effect of nonlinear interactions. As pointed out a long time ago\cite{kivshar_and_peyrard}, modulational instability in discrete systems is a very efficient mechanism to generate discrete solitons. Equation (\ref{eq:1}) admits plane-wave solutions that lead to the nonlinear dispersion relation
\be
\omega(k) = 2 \cos k + (q + 2 g) u_{0}^{2}\ .\label{dr}
\ee
Next, we compute the modulation stability of this plane wave solution by setting
a perturbed solution in the form $u_{n}(t) = [u_{0} + \delta u_{0}(t)] \exp(i k n + \omega t)$.
After replacing in Eq.(\ref{eq:1}) and keeping linear terms in $\delta u_{n}$ only, we obtain the evolution equation for the perturbation,
\begin{eqnarray}
& &i {d\over{dt}} \delta u_{n} + [2 (q+g)u_{0}^{2}-\omega(k)] \delta u_{n} + \nonumber\\
& & (\delta u_{n+1}e^{i k} + \delta_{n-1} \exp^{-i k}) + q u_{0}^{2} \delta u_{n}^{*}+ \\
& &g u_{0}^{2} (\delta u_{n+1}+\delta u_{n-1}+\delta u_{n+1}^{*}+\delta u_{n-1}^{*})=0\ .\nonumber
\end{eqnarray}
We pose $\delta u_{n}(t)$ in the form
\be
\delta u_{n}(t) = u_{1} e^{i(Q n + \Omega t)} + u_{2}^{*} e^{-i(Q n + \Omega^{*} t)},
\ee
leading to the linear system
\begin{eqnarray}
[-\Omega - \omega(k) + a^{+}]\ u_{1} + b\ u_{2} &=&0\nonumber\\
b\ u_{1} + [\Omega - \omega(k) + a^{-}]\ u_{2} &=&0,
\end{eqnarray}
where
$b\equiv q u_{0}^{2} + 2 g u_{0}^{2} \cos Q$ and $a^{\pm}\equiv2 (q+g)u_{0}^2 + 2 \cos(Q\pm k) + 2 g u_{0}^{2} \cos Q$. Nontrivial solution exists provided
\be
\Omega={1\over{2}}\left[ d \pm
\sqrt{d^{2} - 4 b^{2} + 4(\omega(k)-a^{+})(\omega(k)-a^{-})}\right],
\label{eq:Omega}
\ee
where $d\equiv a^{+}-a^{-}$. From this, we define the instability gain
$G\equiv {\mbox{Im}}[\Omega]$. If $G\neq 0$, the nonlinear plane wave experiences
``modulationally instability'' (MI).

\begin{figure}[b]
\includegraphics[width=8.5cm]{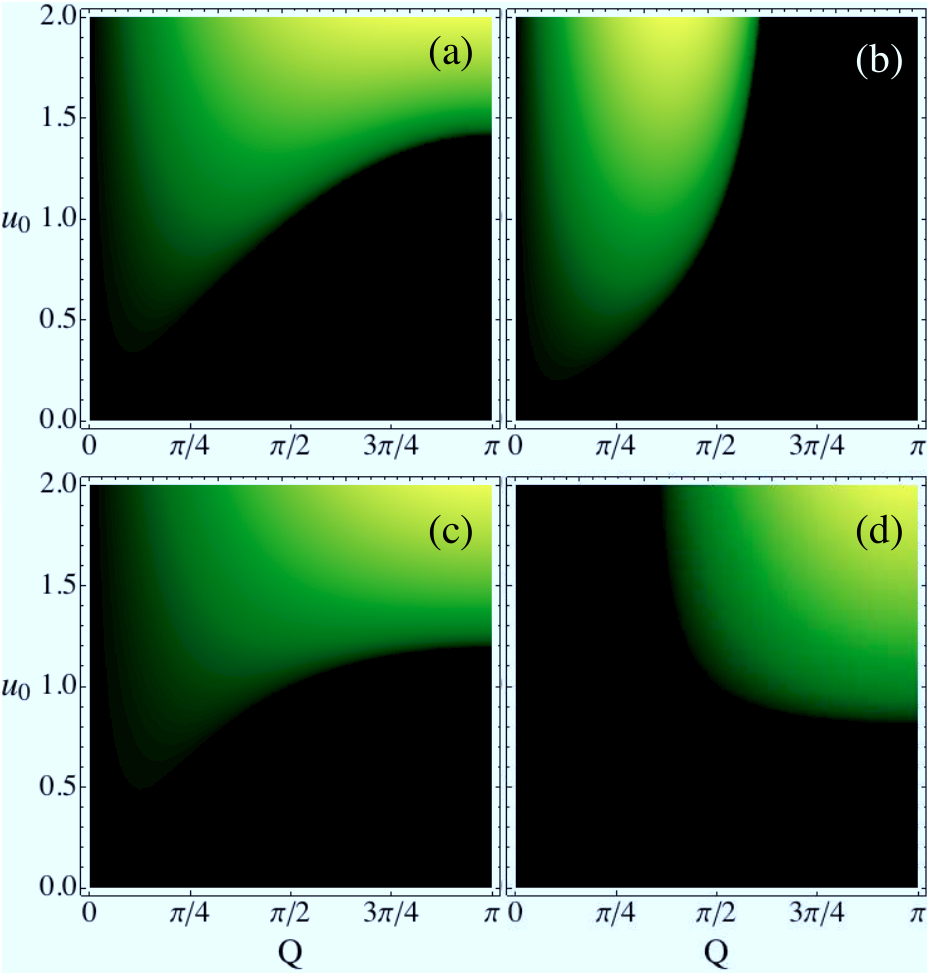}
\caption{(Color online) $G$ for nonlinear plane wave solution for $q=1$ and $g=0$ (a), $g=1$ (b), $g=-0.2$ (c), $g=-1$ (d). Darker (lighter) color means a stable (unstable) region.}
\label{MI}
\end{figure}

Let us discuss the stability of uniform plane waves ($k=0$). In the special case $g=0$, the system reduces to a pure DNLS chain~\cite{MIcubic} and the gain for the  plane wave becomes $G=\pm {\mbox{Im}}[\sqrt{2 \sin(Q/2)^{2}-q u_{0}^{2}}]$, as shown in Fig.\ref{MI}(a), where $G$ is shown in the form of a density plot as a function of $u_{0}$ and $Q$. For $0<q u_{0}^{2} < 2 $, there is MI for $Q<2 \arcsin(\sqrt{q u_{0}^{2}/2})$.
For $q u_{0}^{2}>2$ we have MI for all $Q$ values. Therefore, plane matter waves will be always unstable. 
%
In the case of zero effective mean field nonlinearity ($q=0$) the gain is
given by $G = 4\ {\mbox{Im}}\{[(\cos Q-1)(-1+(1+2 g u_{0}^{2})\cos Q)]^{1/2}\}$.
Analysis of this expression shows that, for $g>0$ and a given $u_{0}$, there is always a $Q$-interval where $G\neq 0$, and the uniform plane wave is always unstable.
On the contrary, for $g<0$, the plane wave is stable for $|u_{0}|^{2}<1/|g|$.
For $q, g>0$, it can be proven from Eq.(\ref{eq:Omega}) that $G\neq 0$ for any $u_{0}$ and, therefore, the plane wave is always unstable [see Fig.~\ref{MI}(b)].
For $q>0, g<0$, it can be shown that for $|g|<q/2$, $G\neq 0$, while for $|g|>q/2$, $G=0$ for $u_{0}^{2}<2/(q+2|g|)$ [see Figs.~\ref{MI}(c) and (d)].

The above results suggest that formation of discrete solitons will be likely in most cases, with the exception of strong attractive dipolar interactions, where a minimum norm will be required. These rough predictions will be confirmed in Section V.

\section{Dimer approach}
\label{dimer}

\begin{figure}[t]
\includegraphics[width=8.5cm]{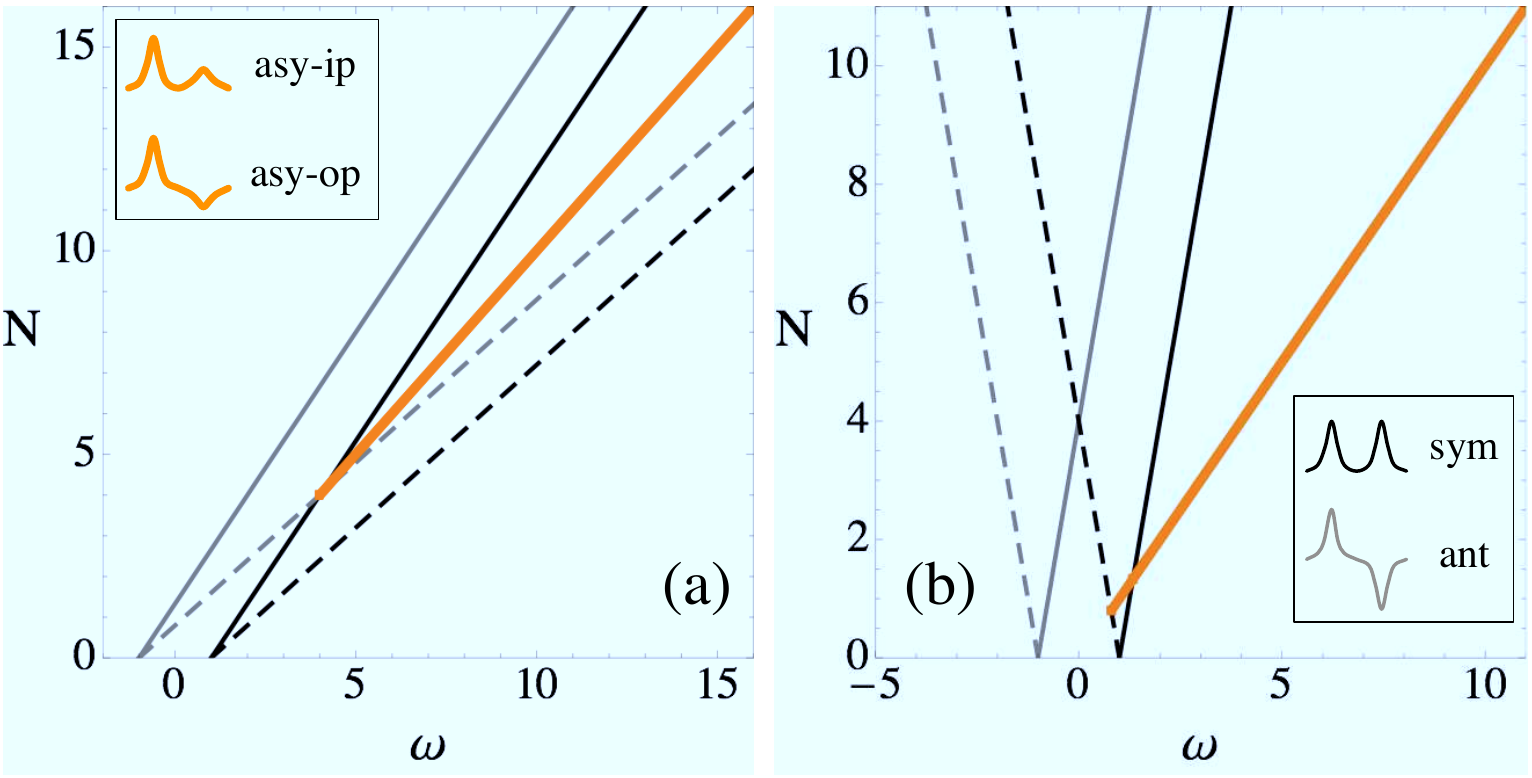}
\caption{(Color online) Norm versus frequency diagrams for $q=1$. Black and gray thin lines correspond to the symmetric and antisymmetric solutions, respectively. (a) $g=0.5$ (continuous) and $g=1.5$ (dashed). (b) $g=-0.5$ (continuous) and $g=-1.5$ (dashed). The thick line represents the asymmetric solution. Insets depict stationary dimer profiles.}
\label{dim1}
\end{figure}

In order to get a deeper understanding of the present model, it will prove useful to consider first the dimer limit $M=2$~\cite{3}. This constitutes an integrable system which can give us some insights about the general phenomenology occurring in larger systems. In particular, it will prove useful when we consider localized surface modes. We solve Eq.(\ref{esta}) for $M=2$
%
\begin{subequations}\label{eq:dimer}
 \begin{align}
\omega u_1 = u_2+qu_1^3+gu_2^2u_1\ , \\
\omega u_2 = u_1+qu_2^3+gu_1^2u_2 \ .
\end{align}
\end{subequations}
%
As a general ansatz, we consider $u_1=A$ and $u_2=\beta A$. After inserting this ansatz in (\ref{eq:dimer}), we obtain three stationary solutions:
\[
\beta=\pm 1\ \ \ \ \text{and}\ \ \ \ \beta = \frac{1}{(q-g)A^2} \ .
\]
The solution $\beta=1$ corresponds to a ``symmetric'' solution ($u_1=u_2$) for which $N_{sym}=2(\omega-1)/(q+g)$. This solution bifurcates from the symmetric linear mode at $\omega=1$. A second solution, $\beta=-1$, corresponds to the ``antisymmetric'' mode ($u_1=-u_2$) with $N_{ant}=2(\omega+1)/(q+g)$, bifurcating from the antisymmetric linear mode at $\omega=-1$. A third solution is called ``asymmetric'' because, in general, $|u_1|\neq |u_2|$. It appears at $\omega_{min}=2q/|q-g|$, the only frequency where $|u_1|=|u_2|$, bifurcating from the ``symmetric'' (``antisymmetric'') solution if $q>g$ ($q<g$) [see thick orange lines in Fig.\ref{dim1}]. This also implies a change in the solution's topology from ``in-phase (\textit{ip})'' to ``out of phase (\textit{op})''. For this asymmetric mode $N_{asy}=\omega/q$, i.e., this mode does not depend at all on the dipolar interaction. A standard linear stability analysis show that the symmetric solution is stable for $N<2/(q-g)$ [i.e., before the onset of the stable asymmetric solution, for $q>g$] while the antisymmetric mode is stable for $N>2/(g-q)$ [i.e., after the onset of the unstable asymmetric solution, for $q<g$].
By fixing $q$ while increasing $g$, we observe that the slope of the $N$ vs. $\omega$ curves for the symmetric and antisymmetric solutions decreases (for $q=g$ both curves coincide with the asymmetric one). It is important to notice that the symmetric solution increases its stable existence region while $\omega_{min}$ increases as $g\rightarrow q$. For $g>q$, the symmetric solution is always stable; the asymmetric one becomes unstable bifurcating, now, from an always stable antisymmetric solution (for $q>0$). Fig.\ref{dim1}(a) shows an example of this phenomenology, where $\omega_{min}=4$ for $g=0.5$ and $g=1.5$. For the sake of simplicity, we have plotted the asymmetric solution as a ``continuous'' thick orange line for both, the stable and unstable, cases.

We now decrease $g$ from zero while keeping $q$ fixed. We see that when $g\rightarrow -q$, the norm of the symmetric and antisymmetric solutions diverges. Before this happens, the global sign of the nonlinearity ($q+g>0$) and the slope  is positive. However, when $q+g<0$ the effective sign becomes negative and the slope changes completely. The asymmetric solution increases its norm as before, because its slope does not depend at all on the $g$-value; i.e., as soon as $q>0$ this solution possesses a positive slope with $\omega_{min}>0$ for any $g$. See Fig.\ref{dim1}(b) as an example of this phenomenology for $g=-0.5$ and $g=-1.5$. From (\ref{dr}) we can see that, for a larger system, this situation will occur when $g\rightarrow -q/2$, due to the larger number of nearest-neighbors. For $g<-q/2$, the fundamental linear mode ($k=0$) will increase its norm by decreasing its frequency. Of course, this will have consequences for localized solutions bifurcating from the linear band, as we will see below.

\subsection{Effective potential}

\begin{figure}[t]
\includegraphics[width=4.25cm]{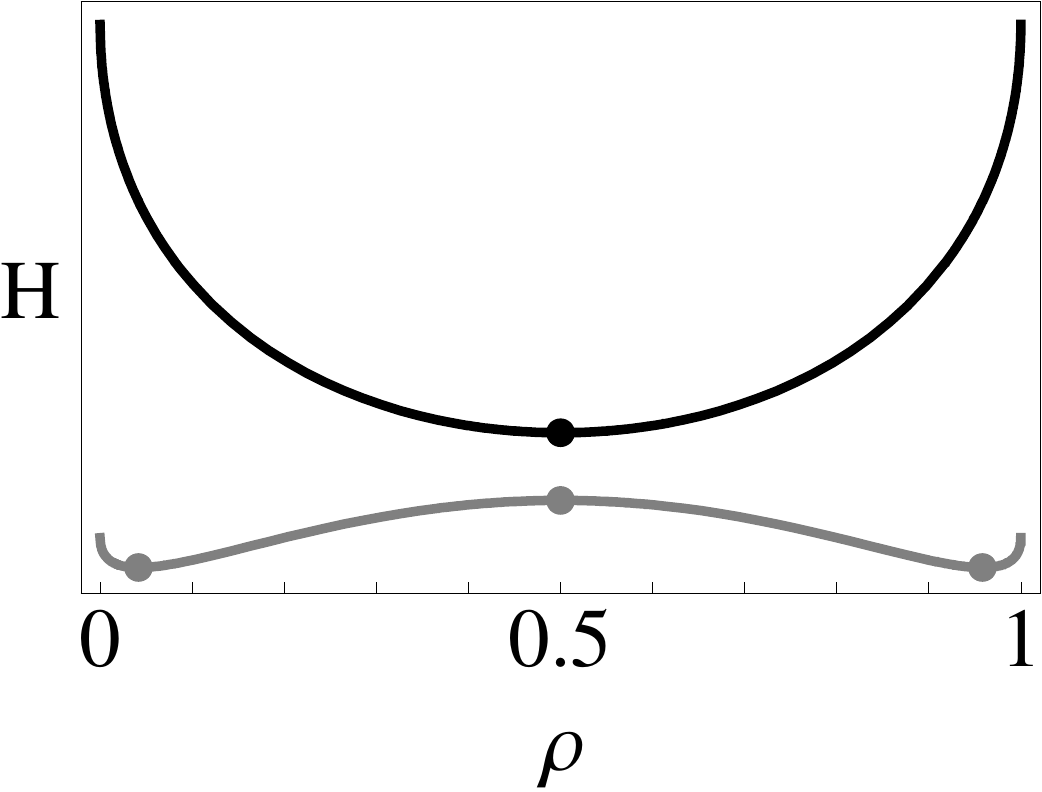}
\includegraphics[width=4.25cm]{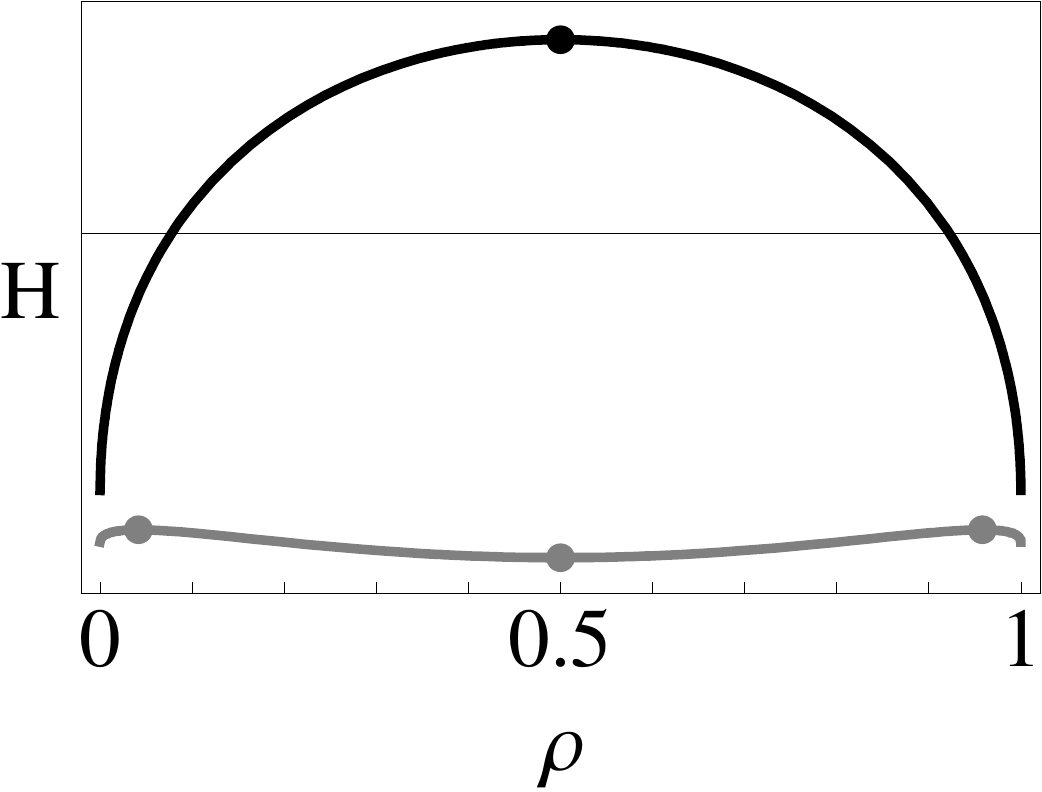}
\caption{Effective potential for $q=1$, $g=0.5$ (left) and $g=1.5$ (right); for $N=1$ (black line) and $N=10$ (gray line). $H$ has been scaled for comparison purposes.}
\label{dim3}
\end{figure}

We now construct an effective potential for the dimer model that connects the stationary solutions found above with a dynamical picture of this problem~\cite{diego}. This will be important to better understand the method we will implement below when dealing with larger systems. For any stationary solution, we can define a center of mass as $\rho\equiv u_2^2/N$, being $N=u_1^2+u_2^2$. $\rho=0$ and $\rho=1$ denote a solution located at site $n=1$ and site $n=2$, respectively; while $\rho=0.5$ implies a solution centered at the inter-site position ($|\beta| =1$). We now express $u_{1}$ and $u_{2}$ in terms of the $\rho$ coordinate: $u_1= \pm\sqrt{N(1-\rho)}$ and $u_2= \pm\sqrt{N\rho)}$. After inserting these expressions in the dimer Hamiltonian, we obtain
\[
H(\rho)=\pm 2N \sqrt{\rho(1-\rho)}-N^2\left[\frac{q}{2}+(g-q)\rho(1-\rho)\right],
\]
where $-$ ($+$) denotes the \textit{ip} (\textit{op}) cases. For a given norm $N$, we look for critical points of the effective potential $H(\rho)$, obtaining
\[
\rho_{sym,ant}=\frac{1}{2}\ \ \ \ \text{and}\ \ \ \ \rho_{asy}=\frac{1}{2}\pm \sqrt{\frac{1}{4}-\frac{1}{N^2 (q-g)^2}}\ ,
\]
as critical points. Therefore, if $N>2/|q-g|$, there are always three stationary solutions: one centered in between the two sites (symmetric or antisymmetric) and two asymmetric solutions with a varying center of mass. For $N_{min}\equiv 2/|q-g|$, the asymmetric solution simply coincides with the symmetric or antisymmetric one [black dots in Fig.\ref{dim3} where $N_{min}=4$]. As the norm increases, one asymmetric solution bifurcates to the left and the other to the right from $\rho=0.5$ [see gray points in Fig.\ref{dim3}]. For $N\gg N_{min}$, the asymmetric solutions locate at $\rho \rightarrow 0,1$.
%
%

The sketched potentials agree perfectly with the properties described before. For $g<q$ and $N<N_{min}$, the only stationary solution is the symmetric one which corresponds to a minima (stable) in the effective potential [see black line in Fig.\ref{dim3}-left]. Asymmetric solutions exist only above some norm threshold and they correspond to two local minima (stable) in the effective potential which, as expected, possesses a local maxima in between corresponding to the unstable symmetric solution. Therefore, the appearance of the asymmetric solution destabilizes the symmetric one. On the other hand, for $g>q$ the situation is quite different. For a small norm the antisymmetric solution is a maximum (unstable) of the effective potential [see black line in Fig.\ref{dim3}-right]. For a larger norm, the asymmetric solution appears as a maximum in the potential $H(\rho)$ being, therefore, unstable while the antisymmetric solution stabilizes. This means that, in this case, the appearance of the asymmetric solution stabilizes the antisymmetric one.

\section{Bulk localized solutions}
\label{su1}

In this section, we will study localized stationary solutions located at the center/middle of the lattice (bulk solitons). Some profiles are shown in the inset of Fig.\ref{dia1} for a lattice of $M=100$ sites. In general, the profiles change in a very defined manner. For a fixed norm we see that, by increasing the value of $g$ from zero, on-site profiles increase its width by increasing the amplitude of the first nearest-neighbors while the inter-site modes does not change significantly. Therefore, a positive dipolar long-range interaction promotes a wider on-site profile due to the increasing relevance of nonlocal amplitudes. For the same reason, the inter-site modes are not effectively affected by this increasing $g$. On the other hand, for a negative value of $g$, solutions become more localized. A decreasing long-range interaction makes the first nearest-neighbor amplitudes smaller, leading to profiles that are very well localized in space.

\subsection{$q,g>0$}

\begin{figure}[b]
\includegraphics[width=8.5cm]{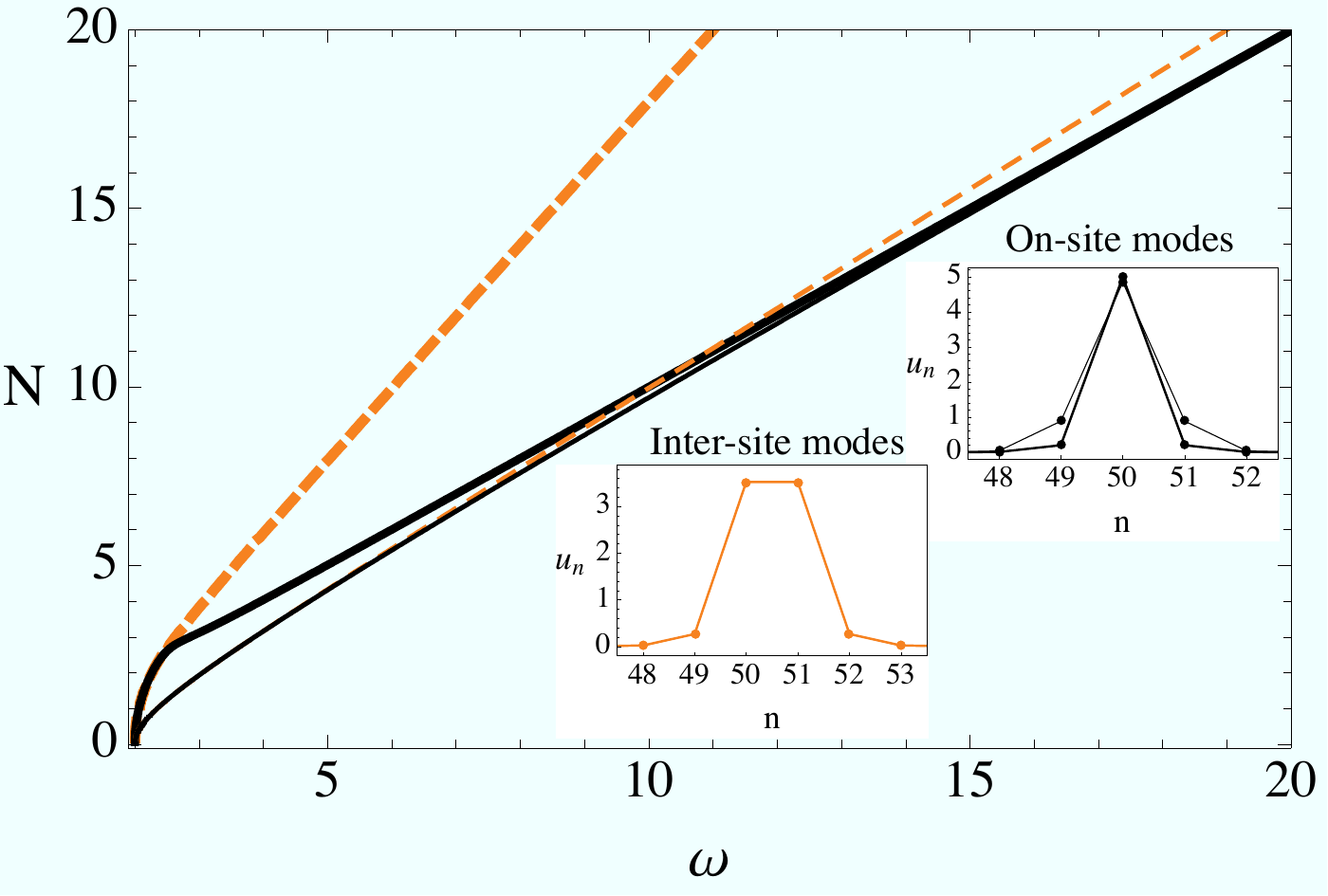}
\caption{(Color online) Norm versus frequency diagrams for $q=1$. Black (orange) curve represents the on-site (inter-site) mode. Thick (thin) lines correspond to $g=0$ ($g=0.8$). Inset: Profiles at $N=25$ for $g=0$ (thick) and $0.8$ (thin).}
\label{dia1}
\end{figure}

First of all, we consider the case where both nonlinear coefficients are positive. We also fix, without loss of generality, $q=1$ and vary $g>0$. For $g=0$, we obtain the well-known DNLS phenomenology where the on-site solution is always stable while the inter-site mode is always unstable [thick lines in Fig.\ref{dia1}].
%
%
If we increase the value of $g$, for instance to $g=0.8$ [thin lines in Fig.\ref{dia1}], we see how the slope of the inter-site family decreases abruptly, getting closer to the one for the on-site solution. (This is somehow similar to the situation encountered for the dimer, where the slope of the symmetric solution approaches the one of the asymmetric solution). In addition, the stability properties start to change for $g$ approaching $q$. At low frequencies, there is a small region where the on-site solution becomes unstable, while the inter-site mode stabilizes. Then, both solutions exchange stability and the previously described behavior is reversed.
%
\begin{figure}[htbp]
\includegraphics[width=8.5cm]{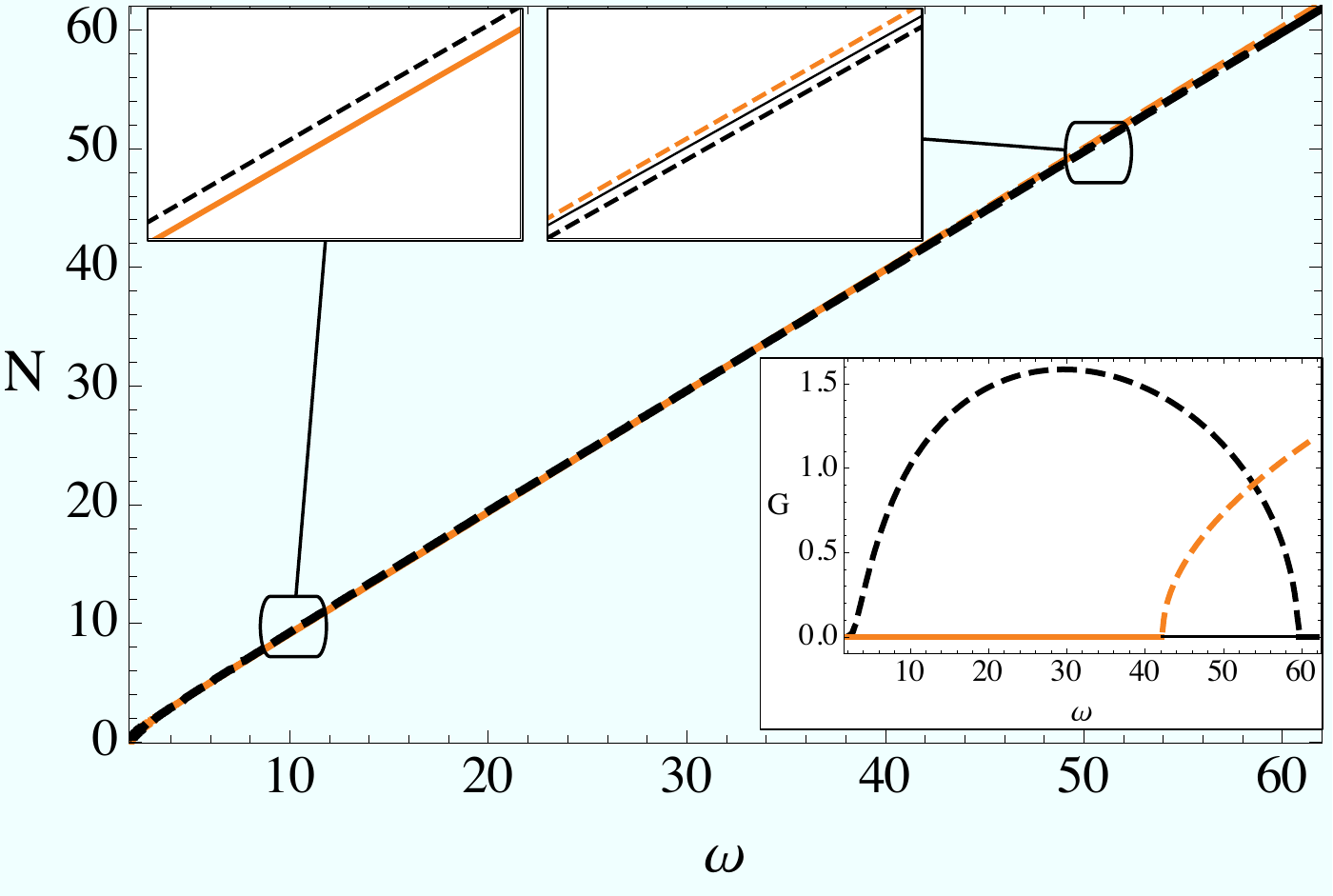}
\caption{(Color online) Norm versus frequency diagram for $q=1$ and $g=0.96$. Black (orange) thick lines correspond to on-site (inter-site) modes, while black thin lines to intermediate solutions. Inset: Instability gain versus frequency.}
\label{dia2}
\end{figure}
%
However, if we increase  the value of $g$, a bit more, for instance to $g=0.96$ [see Fig.\ref{dia2}], we first see - in the $N$ versus $\omega$ diagram - how the solution families approach, being almost undistinguishable from each other. Now, if we take a look of the stability properties, a very interesting phenomenology emerges. The previous apparently trivial exchange of stability is not such. For lower frequencies, as described before, the inter-site solution is stable while the on-site one is unstable; i.e, a completely opposite stability picture compared to the DNLS limit. Then, there is a complete region where both fundamental solutions are simultaneously ``unstable'' [see Fig.\ref{dia2}-inset]. For discrete nonlinear systems this kind of behavior was also observed in more complex models~\cite{mika,Ablowitz,Abd08} which include the same linear and nonlinear dispersion terms plus many others. On the other hand, a completely opposite phenomenology have been predicted for optical saturable systems~\cite{satur,diego} with multiple regions of simultaneously ``stable'' solutions. Therefore, from the fundamental and dynamical point of view, the present phenomenology is very important and, certainly, interesting because, in principle, we could expect good conditions for mobility in one and, also, higher dimensions.

\begin{figure}[htbp]
\includegraphics[width=8.5cm]{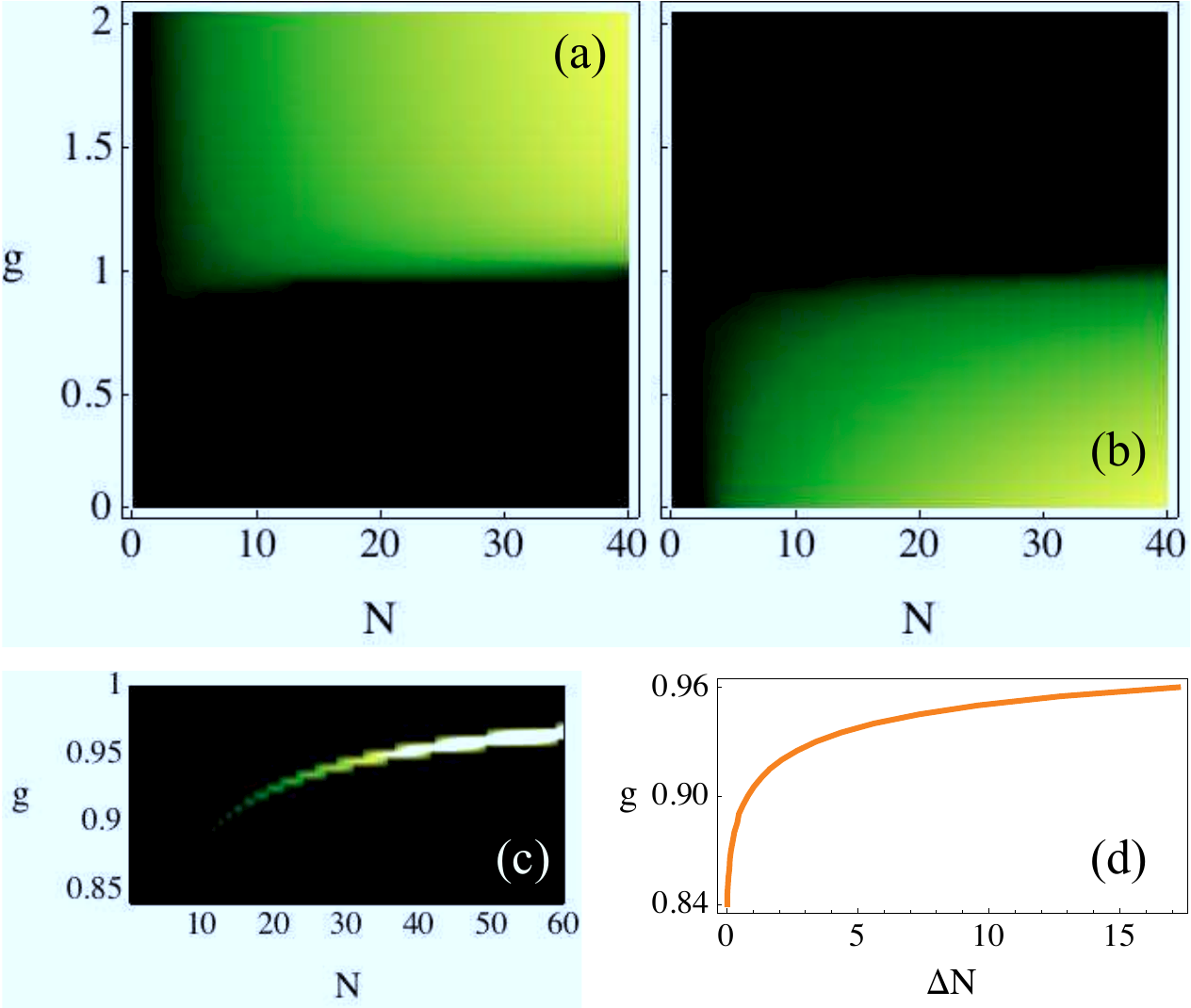}
\caption{(Color online) Instability gain as a function of $g$ and $N$ for (a) on-site and (b) inter-site solutions. (c) Bi-unstable region. (d) $g$ versus $\Delta N$ for the bi-unstable region. ($q=1$).}
\label{bista}
\end{figure}

\begin{figure}[htbp]
\includegraphics[width=8.5cm]{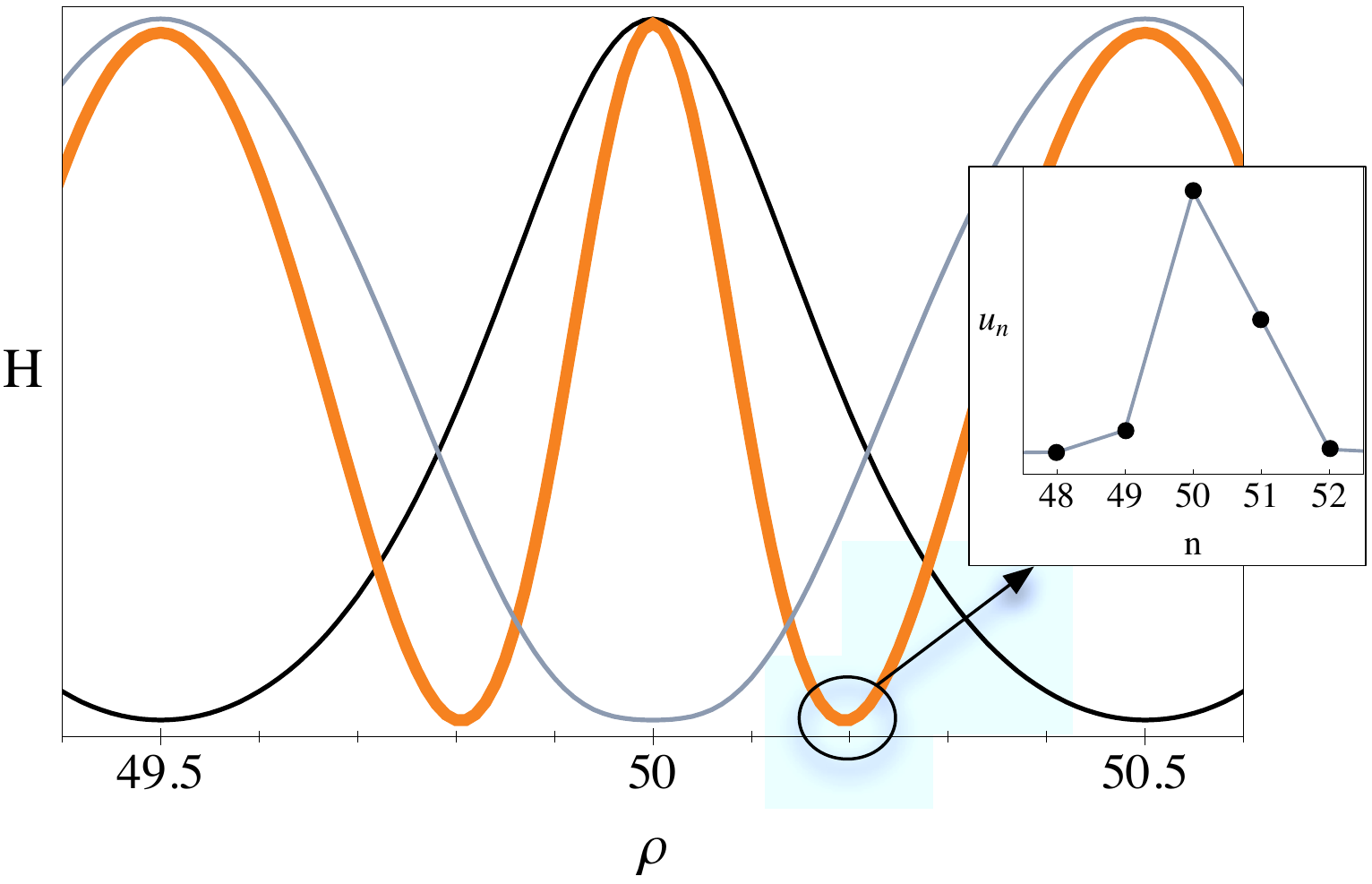}
\caption{(Color online) Hamiltonian versus center of mass por $N=35$ (black line), $N=48$ (orange thick line) and $N=59$ (gray thin line) for $q=1$ and $g=0.96$. Inset: Intermediate solution for $N=48$.}
\label{pots}
\end{figure}

In order to delve deeper in this analysis, we study in detail the stability of both fundamental solutions [see Figs.\ref{bista}(a) and (b) where a brighter (darker) color corresponds to a more unstable (stable) solution]. First of all, from this analysis it is clear that the on-site (inter-site) solution is stable (unstable) if $g<q$, while for $g>q$ the situation is the opposite. Fig.\ref{bista}(c) shows the region, in parameter-space, where both solutions are ``simultaneously'' unstable. As $g$ increases the exchange region also increases showing that, by tuning the dipolar interaction, we could observe this phenomenology for different values of the norm. We computed the width ($\Delta N$) of this bi-unstable region [see Fig.\ref{bista}(d)] for different values of $g$. We can see that $\Delta N$ tends to diverge when $g\rightarrow q$, indicating the full stability exchange for $g>q$.
%
%
From a dynamical point of view, theory tells us that once we have two unstable solutions, both should correspond to a local maxima in a Hamiltonian representation. Therefore, there should exist another solution in between corresponding to a local minima, a stable ``intermediate'' solution (IS) [see inset in Fig.\ref{pots}]. The IS is an asymmetric stationary solution with a profile that varies from an inter-site mode (smaller norm) to an on-site mode (larger norm). In that sense, the IS is a kind of stability carrier that exchanges the stability of both fundamental solutions. Now, we compute an effective potential, i.e. the Hamiltonian ($H$) versus the center of mass ($\rho\equiv \sum_n n |u_n|^2/N$) of different solutions across the lattice, by using a constraint method~\cite{surface,satur} similar to the computation performed for the dimer. (An on-site solution possesses an integer $\rho$-value while the inter-site mode a semi-integer one). In this picture, a stable solution will correspond to a local minima while an unstable solution coincide with a local maxima. For example, in a DNLS phenomenology the on-site solution corresponds to a minima, while the inter-site one to a maxima [see the gray thin curve in Fig.\ref{pots} as an example of this behavior]. Therefore, the effective potential in cubic systems is periodic, and the energy differences between these two fundamental solutions constitute the energy barrier in the system (also known as the Peirls-Nabarro barrier). Mobility will only be possible if the solution has enough kinetic energy to overcome this self-induced energy barrier~\cite{yuca}. However, in more complex systems, the situation is not that simple. For instance, in the present model, the effective energy barriers will depend on the particular value of the norm, $g$, and the particular stability region. Fig.\ref{pots} shows different effective potentials by fixing the value of $g=0.96$ and varying $N$, where we have plotted ``normalized'' $H$-values in order to compare the different cases. First, we can see that for a smaller norm (black line), the inter-site solution corresponds to a minima (stable) while the on-site mode corresponds to a maxima (unstable). In the region where both fundamental solutions are simultaneously unstable (thick orange line), both correspond to a maxima. The stable IS corresponds to a minima located in between both fundamental solutions [see inset in Fig.\ref{pots}]. The potential is again periodic but, now, it has a more complicated ``binary'' geometry. The potential is softer when going from the IS to the inter-site one than when going to the on-site mode. The shape of this potential is also a consequence of the stability properties for both fundamental solutions. For this value of the norm, the on-site solution is more unstable than the inter-site one [see Fig.\ref{dia2}-inset], what coincides with the shape of the two maxima. From this picture, we could predict that an on-site solution will require a smaller kinetic energy to move than the inter-site mode. By increasing the norm, the stability analysis predicts that the on-site (inter-site) mode becomes stable (unstable). The effective potential becomes a simple cubic-like potential as in the DNLS limit.
%
%

In order to corroborate the potential's picture, we integrate numerically model (\ref{eq:1}) to study the dynamics for $N=48$. We initialize our numerical integration with the three stationary solutions of this problem including a kinetic energy with a `kicked' mode of the form: $u_n(0)=u_n \exp{[i k(n-n_c)]}$.
%
\begin{figure}[htbp]
\includegraphics[width=8.5cm]{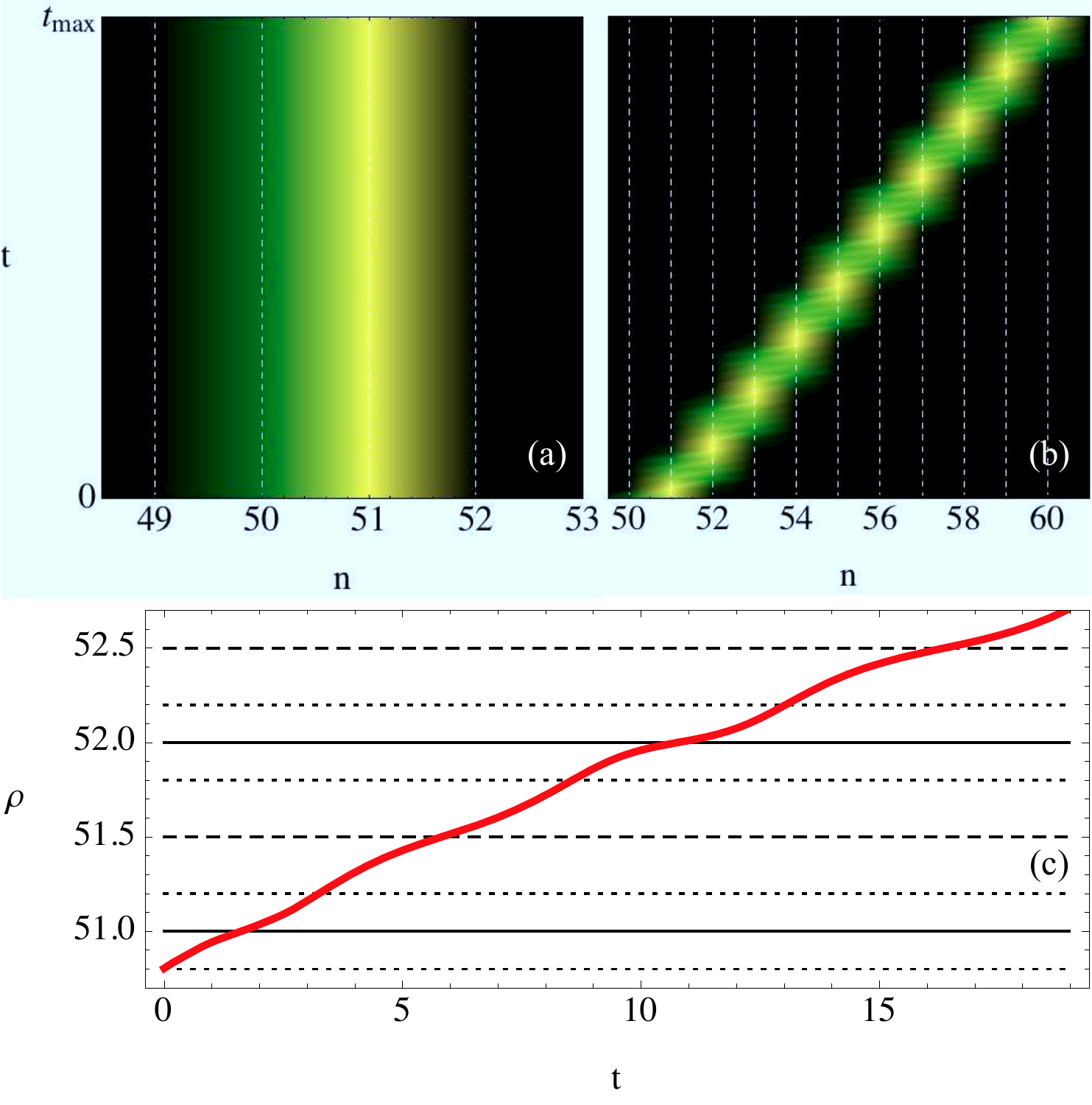}
\caption{(Color online) (a) and (b) Evolution of $|u_n|^2$ of an initial IS ($N=48$) for $k=0$ and for $k=0.18$, respectively, for $t_{max}=100$. (c) Center of mass versus time for $k=0.18$.}
\label{dina1}
\end{figure}
%
First of all, we observe a stable propagation of the intermediate solution (IS) [see Fig.\ref{dina1}(a)], as expected from the computed potential and the stability analysis. Contrary to what is typically expected, this asymmetric solution propagates stably for long times. From Fig.\ref{pots} we know that some energy should be added to the IS in order to set it in motion. Fig.\ref{dina1}(b) shows an example of very good mobility with no visible radiation from tails; i.e., a coherent movement of the atomic wave-packet. The solution just propagates ``feeling'' the topology of its own potential. Fig.\ref{dina1}(c) shows how the velocity (slope) of the propagating solution changes according to the energy surface. When it passes through integer (thin horizontal lines) or semi-integer (dashed horizontal lines) positions, the velocity decreases because these points correspond to local maxima. On the other hand, when it passes through positions related to IS (dotted horizontal lines), the velocity increases, as expected from general dynamical arguments.
%
\begin{figure}[htbp]
\includegraphics[width=8.5cm]{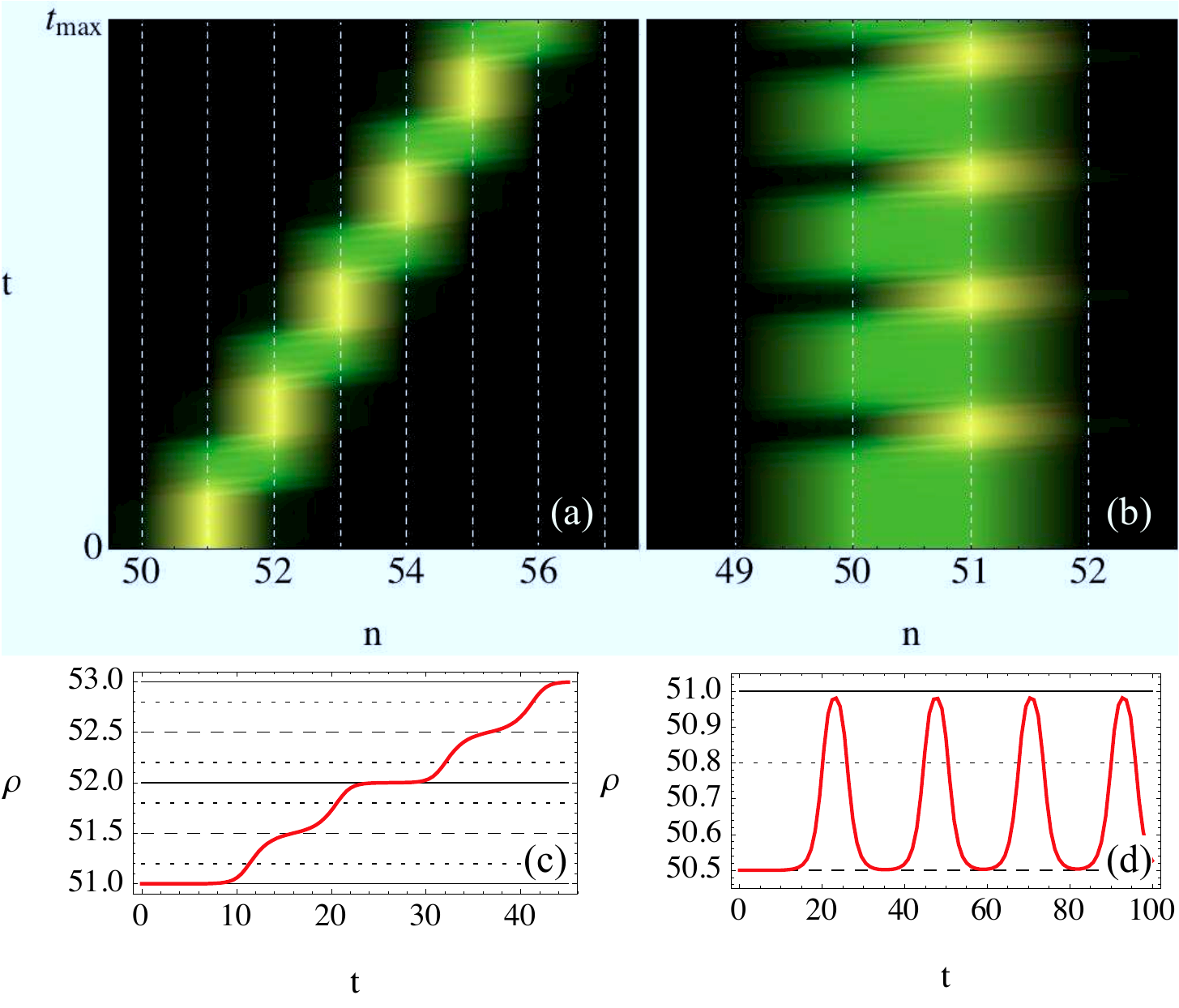}
\caption{(Color online) (a) and (b) Evolution of $|u_n|^2$ for an initial on-site and inter-site solution, respectively, for $N=48$. (c) and (d) Center of mass versus time for cases (a) and (b). $k=0$ and $t_{max}=100$.}
\label{dina2}
\end{figure}
%

Now, we initialize our numerical simulation with unstable on-site and inter-site solutions at $N=48$. It is important to notice that, in this case, $H_{onsite}>H_{intersite}$ [see Fig.\ref{pots}]. The on-site solution is located in a very sharp local maximum, so a very unstable dynamics is expected. In addition, as this maximum is larger than the one for the inter-site solution, a good mobility is expected through the lattice, if radiation from tails is not too large. Figs.\ref{dina2}(a) and (c) show the propagation of an on-site solution without any initial kinetic energy. Dynamics corroborate the expected phenomenology from the potential analysis: the velocity tends to zero when the solution passes through an on-site position ($\rho=51$, $52$ and $53$, in this example); the velocity is small but not zero when solution crosses an inter-site position ($\rho=51.5$ and $52.5$, in this example); and the velocity increases to a maximum when the solution passes through an IS position ($\rho=51.2$, $51.8$, $52.2$ and $52.8$, in this example). On the other hand, by initializing the numerics with an unstable inter-site solution, without any initial kinetic energy, the dynamics is less unstable than before. The solution takes more time to destabilize and to start an oscillatory dynamics inside the potential well [see Figs.\ref{dina2}(b) and (d)]. The profile changes from an inter-site mode (zero velocity) to an IS (maximum velocity) and then it approaches the on-site solution (zero velocity). Then, the solution goes back and the cycle starts again. This system could be used as an ``atomic clock'', where the fundamental period would depend only on the amount of particles in the system ($N$).

For $g>q$, the inter-site modes become always stable while the on-site solutions are just unstable. This phenomenology is completely the opposite to the DNLS one, therefore when the dipolar interaction is larger than the mean field nonlinearity broader solutions are favored.

\subsection{$q>0$, $g<0$}

\begin{figure}[htbp]
\includegraphics[width=8.5cm]{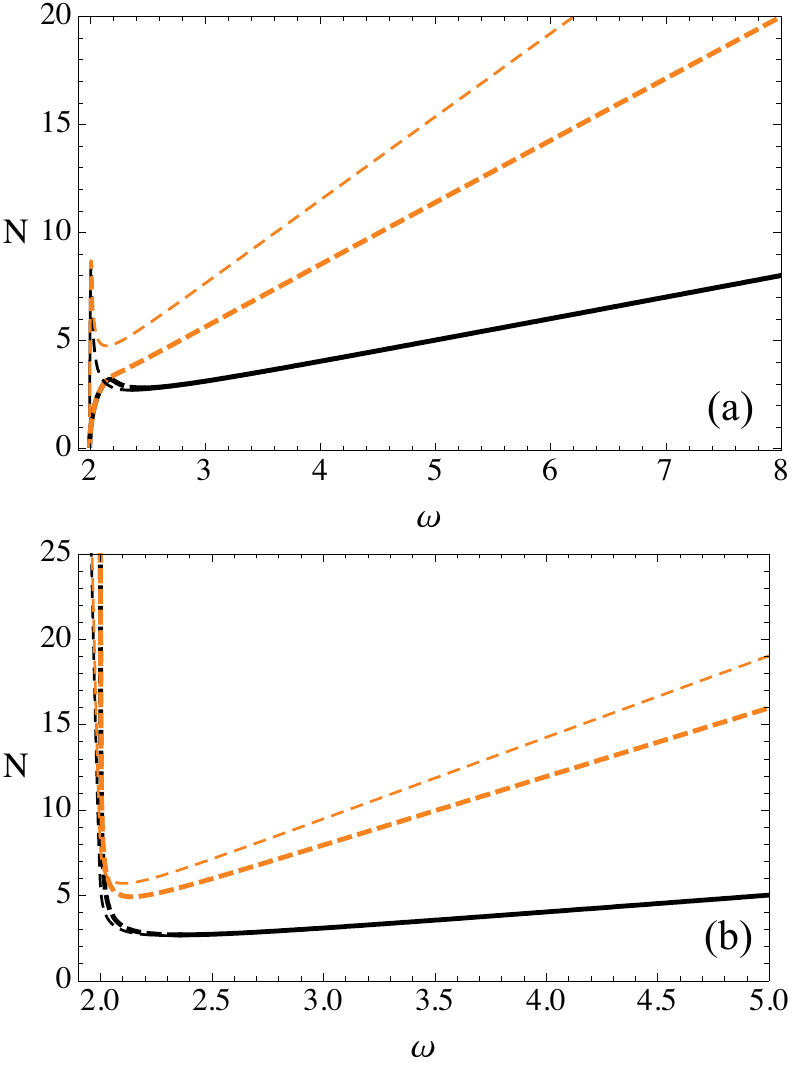}
\caption{(Color online) Norm versus frequency diagrams for on-site (black) and inter-site (orange) modes. (a) $g=-0.3$ (thick) and $g=-0.48$ (thin). (b) $g=-0.5$ (thick) and $g=-0.58$ (thin). ($q=1$).}
\label{dia3}
\end{figure}

Now, we consider $q=1$ and $g<0$ for which the inter-site mode is always unstable in the whole range of parameters. For $g\lesssim -0.3$, the on-site solution presents a critical norm, where the slope changes its curvature and the solution destabilizes [see thick lines in Fig.\ref{dia3}(a)]. Then, it fuses with the inter-site mode becoming stable when changing again its curvature (similar to the  case of a 2D DNLS model). After this, both modes just decrease their norm up to the edge of the linear band at $\omega=2$.
%
%
By decreasing the value of $g$ further, we start to observe that the inter-site solution also shows a change of curvature for a given norm [see as an example thin lines in Fig.\ref{dia3}(a)]. Moreover, both fundamental solutions still bifurcate from the linear band ($\omega\rightarrow 2$) with a large slope as $g$ approaches the critical value of $g=-0.5$.
When $g=-0.5$ and, therefore, $q+2g=0$ [see thick lines in Fig.\ref{dia3}(b)], both solutions increase their norm to infinite at a given minimum norm value. The decreasing branch coming from $\omega\gg 2$ should connect with a family which bifurcate from the linear band ($\omega\sim 2$) at an infinite rate, as predicted from Eq.(\ref{dr}). Therefore, a norm threshold should appear in between in order to connect the two family branches.
If we decrease the value of $g$ even further [see thin lines in Fig.\ref{dia3}(b)], we observe that the solutions crosses the edge and enter into the linear band. There is a branch that originates at $\omega= 2$ which increases with negative slope as Eq.(\ref{dr}) predicts for the fundamental mode when $q+2g<0$. (This is similar to the behavior observed for the dimer in Fig.\ref{dim1}(b). In that case, the symmetric mode starts to increase its norm in the negative direction of frequencies, identical to the behavior observed for bulk solutions originating from the linear band).
For even more negative $g$-values, solutions start to deviate from each other with very different norm thresholds. The inter-site solution is always unstable and its slope and norm threshold increase as $g$ decreases. In fact, for $g\leqslant -1$ no inter-site solution was numerically found. On the other hand, the on-site solution does not feel the decreasing value of $g$, keeping its slope constant and being stable for all norms above norm threshold.

\section{Analytical estimates}

In order to obtain a deeper understanding of our numerical findings, we test  the expected phenomenology of the stationary model (\ref{esta}) in two limit regimes: large and small norm. Let us consider first the large norm limit. In general, an on-site localized profile will always have a maximum amplitude $A$ while its nearest-neighbor sites will have a value $\beta A$, with $|\beta|<1$. On the other hand, an inter-site solution will always have two main peaks ``$B$'' and two symmetric neighboring  sites with amplitude $\epsilon B$ ($|\epsilon|<1$). By inserting these {\em ans\"{a}tze} in the equation for the center site, we get $\omega_{on-site}=2\beta+(q+2g\beta^2)A^2$ and $\omega_{inter-site}=(1+\epsilon)+[q+g(1+\epsilon^2)]B^2$. It is known that for large norm, solutions are, in general, highly  localized, i.e. $\beta,\epsilon\rightarrow 0$. In addition, for an on-site mode the norm is essentially given  by $A^2$ while for the inter-site mode is approximately $2B^2$. Therefore, to first approximation, $N_{on-site}^{large}\sim \omega/q$ and $N_{inter-site}^{large}\sim 2\omega/(q+g)$. Thus, the increment in norm for a large frequency is linear and independent of the $g$-value for the on-site mode. For the inter-site mode, there is an special case, $q=g$, where the two fundamental solutions coincide, with value $N\sim \omega/q$. Therefore, for $g<q$, the inter-site solution has a larger norm for the same frequency, being therefore unstable. However, for $g>q$ the on-site solution has a larger norm for the same frequency, being now unstable. In that sense, our estimate predicts a change in the stability properties for the fundamental solutions for $q\approx g$. On the other hand, for $g<0$, our estimates predict that the inter-site slope increases while the on-site one keeps equal. As an example, for $g=-0.5$ the inter-site norm is around four times larger than the on-site norm. If $g\rightarrow -1$, our estimates say that the inter-site solution should diverge (disappearing for $g<-1$) while the on-site mode remain unaltered. All these analytical predictions are in perfect agreement with our numerical results shown in section \ref{su1}.

Let us take now the small norm limit. Typically, we know that when norm decreases solutions become wider, bifurcating from the linear band fundamental mode $k=0$, if $q>0$ (or $k=\pi$ if $q<0$). Thus, in this limit, we can assume that $\beta,\epsilon\rightarrow 1$ (meaning that all sites are approximately equally excited) and $N^{small}\sim M (\omega-2)/(q+2g)$, for both fundamental solutions. This expression tells us that the initial slope is larger when the system is larger and that the initial frequency for this solution is $\omega=2$. These estimates also agree with our numerical findings, including the one which predicts that, for $q+2g=0$, the solutions bifurcating from the linear band will diverge, generating a norm threshold that is independent of the lattice size. For $q+2g<0$ our estimate suggests that $\omega$ should be smaller than $2$ in order to keep the norm positive. Therefore, the solution bifurcating from the linear band increases its norm by initially decreasing its frequency.

There is another issue related to a change of topology for the on-site solution. In the dimer problem, we saw that for $g<q$, the asymmetric solution was unstaggered ($u_1 u_2>0$), same as the symmetric solution from where it bifurcates. However, for $g>q$, the asymmetric solution changed its topology becoming staggered ($u_1 u_2<0$), same as the antisymmetric solution from where it bifurcates. Let us try to predict what would happen for bulk solutions. We can do an analytic approximation of the profile by using ``strongly localized modes (SLM)''. We consider the following stationary ansatz for the on-site solution $u_n^{on}\approx A\{0,...,0,\beta,1,\beta,0,...,0\}$. By replacing these expressions in (\ref{esta}), for the center and first nearest-neighbor sites, we find that
\[
\beta=\frac{-(q-g) A^2\pm \sqrt{8+(q-g)^2 A^4}}{4}\ .
\]
For $q>g$, the ``$+$'' sign applies, being $\beta>0$ and the solution is unstaggered ($u_n u_{n+1}>0\ \forall n$). In addition, as $g$ approaches $q$, $\beta$ increases and the on-site solution becomes wider. For $q<g$ the correct sign is ``$-$'', implying that $\beta<0$ and that the solution losses its topology. Therefore, similar to the dimer case, there is a change in the topology of solutions at least for $q\lesssim g$, where $\omega$ increases monotonically with the increment of the norm. We numerically found different on-site solutions for $q<g$. We continue the on-site unstaggered's family, however the first nearest-neighbor amplitudes becomes very large and the SLM approach does not describe the solutions properly. On the other hand, we also found on-site solutions where the sign of the main peak was different to the sign of the first nearest-neighbor amplitudes, coinciding with our prediction. However, this solution does not belong to the family of fundamental solutions we focus on this work. Finally, for $g<0$ (and $q>0$), the on-site solutions are always unstaggered ($\beta>0$) being more and more localized as $g$ decreases.

For the inter-site solution we use the stationary ansatz $u_n^{in}\approx B\{0,...,0,\epsilon,1,1,\epsilon,0,...,0\}$, obtaining
\[
\epsilon=\frac{-(1+q B^2)+ \sqrt{4+(1+q B^2)^2}}{2}\ ,
\]
which displays no dependence on the $g$-value. That means that - for example - for $q=0$ this solution will not exist in the large norm limit (this is not the case of the on-site solution). Therefore, no topological transitions are expected for this type of solutions being always $\epsilon>0$, for all $q, g$.

\section{Surface localized solutions}

\begin{figure}[htbp]
\includegraphics[width=8.5cm]{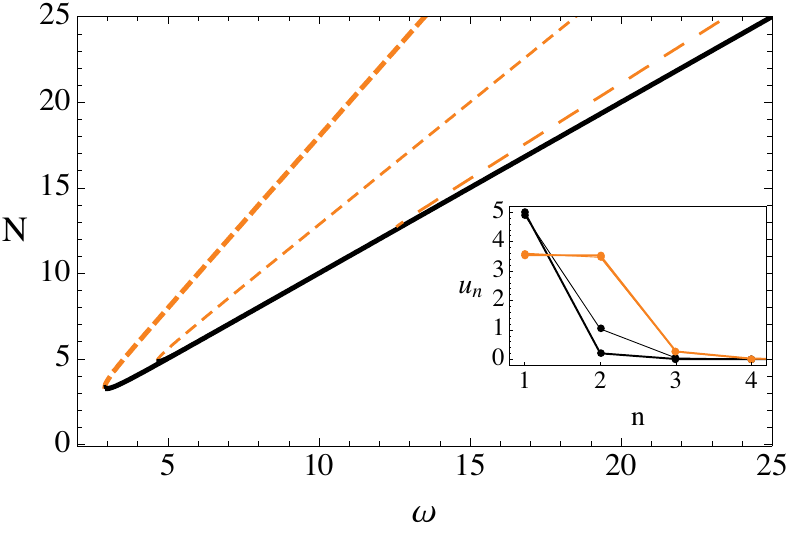}
\caption{(Color online) Norm versus frequency diagrams for $q=1$. Black (orange) curve represents the on-site (inter-site) modes. Thick lines corresponds to $g=0$. Dashed and long-dashed thin lines correspond to $g=0.4$ and $g=0.8$, respectively. Inset: Profiles for $N=25$ and for $g=0$ (thick) and $0.8$ (thin).}
\label{sur1}
\end{figure}

We consider now fundamental, on-site and inter-site, solutions located at the right surface of a 1D lattice, i.e. ``surface solutions'', as shown in Fig.\ref{sur1}-inset. This kind of localized modes possesses a norm threshold for their excitation in usual DNLS lattices~\cite{surface}. Therefore, it turns interesting to study the effect of the dipolar interaction in the excitation of such structures.

The main phenomenology that we observe concerning norm thresholds, consists of their increment as $g$ grows from zero. Therefore, it is more difficult to sustain a localized solution at the surface if the long-range dipolar interactions increase. The increasing repulsive character of the surface become evident from the norm versus frequency diagrams [see Fig.\ref{sur1}]. In all these curves both - on-site and inter-site - solutions get connected after norm threshold where the on-site solution changes its curvature and fuses with the inter-site mode. Therefore, when $g\rightarrow q$, and the norm threshold tends to infinite, both solutions disappear. In this case, this happens when the inter-site solution decreases its slope and approaches the one for the on-site solutions, as Fig.\ref{sur1} shows. As a strong consequence, {\em unstaggered on-site and inter-site surface solutions do not exist for $g>q$}.
%
%

\begin{figure}[t]
\includegraphics[width=8.5cm]{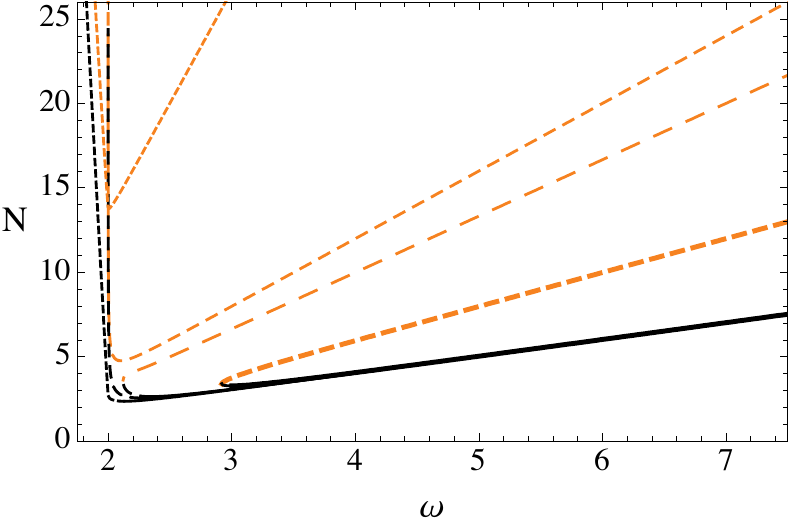}
\caption{(Color online) Norm versus frequency diagrams for $q=1$. Black (orange) curve represents the on-site (inter-site) modes. Thick lines corresponds to $g=0$. Long, middle, and short dashed thin lines correspond to $g=-0.4$, $g=-0.5$ and $g=-0.85$, respectively.}
\label{sur2}
\end{figure}

On the other hand, when $g$ is lower than zero the norm threshold decreases. Fig.\ref{sur2} shows this behavior for $g=-0.4$ where both solutions connect each other as in the DNLS limit. However, for smaller $g$-values both solutions posses a norm/frequency threshold increasing their norm for lower frequencies. If $g\geqslant-0.5$, both solutions connect each other when approaching the linear band edge. For $g=-0.5$ both solutions tend to the linear band edge increasing their norm to infinite at $\omega\rightarrow 2$ [see Fig.\ref{sur2}]. As for the bulk solutions, the surface ones bifurcating from the linear band increase their norm because their minimal frequency decreases for $g<-q/2$ (the effective ``linear-band-border'' also decreases). Fig.\ref{sur2} shows how, for $g=-0.85$, solutions touch the linear-band-border at $\omega=2$ but at very different norm values. As we analytically predicted for inter-site bulk solutions, for surface ones the norm also diverges when $g\rightarrow-q$. Therefore, the slope of this family increases very rapidly in comparison to the on-site one [see Fig.\ref{sur2}]. That implies a larger norm threshold for the inter-site mode, which diverges when $g\rightarrow -q$. Fig.\ref{sur2} shows that the large norm limit for the on-site solution does not depend on the $g$-value.

%
%

\begin{figure}[t]
\includegraphics[width=8.5cm]{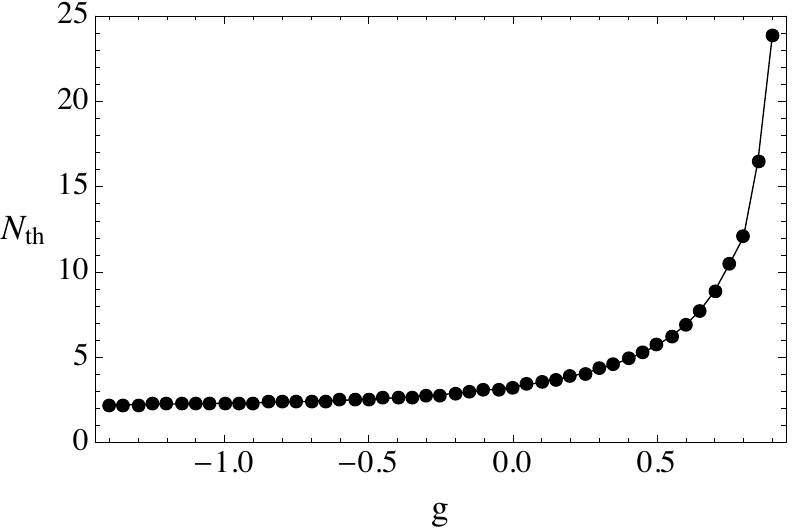}
\caption{Norm threshold versus dipolar interaction strength for $q=1$.}
\label{th}
\end{figure}

As before, on-site surface solutions increase their width while the $g$-values increase [see Fig.\ref{sur1}-inset]. For $g<0$, the opposite is true. On the other hand, inter-site mode profiles are not really affected by the long-range term. In general, our analytical estimates for the large frequency limit of bulk solutions are the same than the ones for the surface solutions. In this limit, on-site solutions are composed essentially by one peak (independent of the particular excited site), while inter-site modes are composed by just two peaks. The SLM approximation for the on-site solution located at the right surface {\em reduces to the dimer model} with the same type of solutions previously described in section \ref{dimer}. Therefore, a similar transition is expected for surface solutions; i.e., the on-site unstaggered surface solution disappears for $g>q$ while its frequency threshold also increases as $g\rightarrow q$, being infinite for $q=g$. For $g<0$, the frequency/norm threshold for the on-site mode decreases as $g$ becomes more negative.
Fig.\ref{th} shows a diagram with the numerically computed minimal norm ($N_{th}$) to excite a localized on-site solution at the surface for different values of $g$. We clearly see the effect of the long-range dipolar interaction: for $g>0$ the norm threshold increases, diverging for $g\rightarrow 1$. This is because the first nearest-neighbor ($u_2$) becomes more and more important and a localized solution requires now a larger norm (nonlinearity) to be trapped at the surface of the optical lattice. However, for $g<0$ the dipolar interaction reduces the effective nonlinearity and, therefore, the frequency of the solution decreases, making possible to excite a solution with smaller norm. These numerical results coincide perfectly with our estimates stemming  from the dimer phenomenology. From the application point of view, a negative dipolar interaction would make possible the excitation and observation of surface states in simpler experimental conditions.
%
%
Phenomenologically speaking, the effective energy potentials in the presence of a boundary look as the ones shown in Ref.\cite{surface}. Above norm thresholds, stationary solutions (extrema) can be excited in all sites of the lattice. However, for a smaller norm, the repulsive effect of the surface is non-negligible, and the nonlinearity is not able to create a localized state at the first or next sites of the lattice.

\section{Conclusions}

In this work we studied the interplay of scattering length and dipolar interactions on nonlinear localized modes (bulk and surface) of Bose-Einstein condensates in deep optical lattices. We started analyzing  the process of modulational instability of nonlinear plane wave in a dipolar nonlinear lattice and established the regions of instability. In particular, for vanishing local atomic scattering length we showed  that, for attractive dipolar interactions the nonlinear plane wave is linearly stable below a threshold amplitude, otherwise they are always unstable. That allowed us to find the favorable conditions for the existence of discrete solitons. Then, using this information, the existence and stability of bulk discrete solitons was investigated analytically and confirmed by numerical simulations. To study the existence and properties of surface discrete solitons we started with an analysis of the dimer configuration, where the properties of symmetric and antisymmetric modes including the stability diagrams and bifurcations was investigated in closed form. For the case of a bulk medium, we analyzed properties of fundamental on-site and inter-site localized modes. In particular we showed that for attractive local and nonlocal nonlinearities, there is a whole region region in parameter space
where both fundamental modes are simultaneously unstable. We also predicted a possible regime where the mode changes periodically from an inter-site mode (zero velocity), to an intermediate state (maximum velocity) and then back to an on-site mode (zero velocity). This system could be used as a precise `atomic clock'. Results for expected shape of profiles agree well with the computed numerical profiles. We also investigated on-site and inter-site modes localized at the surface of 1D lattice, i.e. surface solitons. We found and analyzed the form and properties of surface localized solutions finding a strong dependence on the effect of nonlocal nonlinearities.

The authors acknowledge support from FONDECYT Grants 1080374 and 1110142, and
Programa de Financiamiento Basal de CONICYT (FB0824/2008). F.Kh.A. acknowledges a  Marie Curie Grant No.PIIF-GA-2009-236099(NOMATOS).

\end{document}